\documentclass[%
 reprint,
superscriptaddress,
 amsmath,amssymb,
 aps,
prl,
longbibliography
]{revtex4-1}

\usepackage[normalem]{ulem}%Allows to strike out text using \sout{...} (for editing)
\usepackage{graphicx}% Include figure files
\usepackage{dcolumn}% Align table columns on decimal point
\usepackage{bm}% bold math
\usepackage{bbm}
\usepackage[dvipsnames]{xcolor}
\begin{document}

\title{Theory of spin-polarized current flow through a localized spin triplet state}%: a divacancy in silicon carbide}

%Predicted all-electrical signature of divacancy orientation in silicon carbide and spin polarization efficiency from a spin-polarized contact
\author{Stephen R. McMillan}
\affiliation{Department of Physics and Astronomy, University of Iowa, Iowa City, Iowa 52242, USA}

\author{Michael E. Flatt\'{e}}
\affiliation{Department of Physics and Astronomy, University of Iowa, Iowa City, Iowa 52242, USA}
\affiliation{Department of Applied Physics, Eindhoven University of Technology, Eindhoven, The Netherlands}

\date{\today}

\begin{abstract}
We derive a formalism describing quantum-coherent features of  spin-polarized charge current through a partially-polarized spin triplet defect in a \textit{transverse} magnetic field. 
We
predict distinct few-milli-tesla-dc magnetoresistance signatures that  identify a \textit{single} spin-triplet center's character  and reveal the   orientation of the spin triplet's zero-field splitting axis relative to the magnetic contact's polarization. For example, in 4H-SiC the single \textit{(hh)}, \textit{(kk)}, \textit{(hk)}, and \textit{(kh)}  divacancies are all distinct. 
Spin-polarized current flow  efficiently polarizes the spin, potentially electrically initializing spin-triplet-based qubits. 
\end{abstract}

\maketitle
%%%%%%%%%%%% INTRODUCTION %%%%%%%%%%%%
Spin-based technology relies primarily on the ability to predict and control coherent spin dynamics\cite{Meier1984,Awschalom2002,Hanson2007,Loss1998}. Efforts to control single solid-state spins have been underway for decades, with defects in semiconducting hosts providing robust and tunable realizations of coherent spin centers\cite{Kane1998,Balasubramanian2009,Koenraad2011} that can be applied to nanoscale sensing\cite{Degen2017},  quantum information processing\cite{Nielsen2010} and single photon emission\cite{Aharonovich2016}. The capability to deterministically place and identify defects has advanced in parallel\cite{Schofield2003,Koenraad2011,Fuechsle2012, Schofield2013,Lee2019, Achal2018, Bradac2019,Stock2020} as a key enabling step for large-scale quantum-coherent systems.  
%In addition to deterministic placement, single spin readout techniques allow for conception of single-spin based technology. 
For such goals the neutral divacancies in 4H-SiC possess several advantages: long coherence times\cite{Baranov2005,Koehl2011,Klimov2015,Christle2015,5secSiC},  even at room temperature, optical initialization and readout properties similar to NV$^-$ centers,  emission in the telecom range,\cite{Casas2017,Christle2017} and the  potential to be manipulated  within electrical semiconductor device structures\cite{Anderson2019, Candido2020}. Prior work  used optical techniques to polarize or probe the spin~1 divacancy, followed by manipulation with external fields (including microwave irradiation), however these same operations could in principle be performed entirely  electrically  on far smaller scales than optical wavelengths. Even without coherent microwave manipulation the spin-coherent nature of spin-polarized transport through defects produces remarkable dc magnetoresistive features, whether through coherent\cite{Song2014,Yue2015} or incoherent\cite{Inoue2015,Harmon2021,McMillan2020} orbital transport through the defect. However these dc magnetoresistance theories\cite{Song2014,Yue2015,Inoue2015,Harmon2021,McMillan2020}  apply to defects with a single orbital state, and a spin transition between spin~1/2 and spin~0 ($1/2 \leftrightarrow 0$). 
A SiC divacancy undergoes spin $1\leftrightarrow 1/2$ transitions during electrical transport, and offers distinct features from spin $1/2 \leftrightarrow 0$  transitions, such as  zero-field splitting of the spin~1 state.

%\cite{Anderson1961,Appelbaum1969}

\begin{figure}[htbp!]
\begin{centering}
       \includegraphics[width = \columnwidth, scale=0.4]{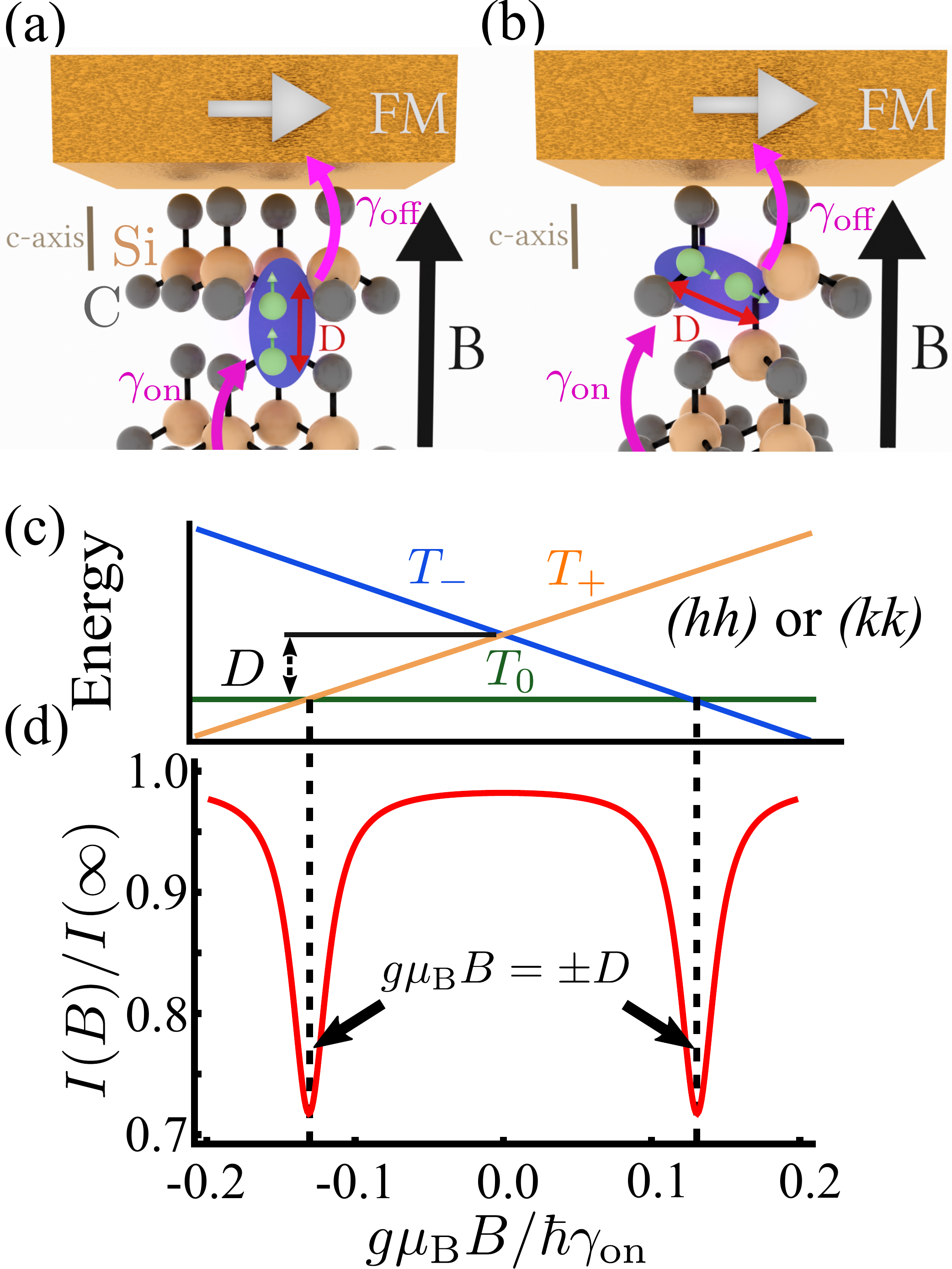}
       \caption[]
{Schematic current path for an electron through a (a) \textit{(kk)} divacancy  or (b) basal \textit{(kh)} divacancy in 4H-SiC. The applied magnetic field {\bf B} $\parallel c$ axis of the crystal and the surface normal. The magnetic contact (FM)  polarization $\perp c$ axis.  The red double-ended arrow  represents the axis of the zero-field splitting (ZFS) $D$. Unpolarized carriers hop to the divacancy from the bulk with rate $\gamma_\text{N}$ and from the divacancy to the FM  with rate $\gamma_\text{F}$. (c) Energy eigenstates for the  divacancy in (a), showing the ZFS and degeneracies for $g\mu_B B = \pm D$. (d) inverse MR for (a) with  $\gamma_\text{F}/\gamma_\text{N}=0.02$. Degeneracies in (c) produce current drops in (d). 
}\label{fig:SetupC3vMR}
       \end{centering}
\end{figure}

Here we calculate the dc magnetoresistance (MR) of spin-polarized current through a localized spin-triplet state such as a  divacancy in 4H-SiC. Our formalism for this dynamical process  
tracks the  spin~1 state (neutral divacancy) with a $4\times 4$ density matrix $\rho_1$, which must be projected onto the triplet subspace, and  the spin~1/2 state (ionized divacancy) with a $2\times 2$ density matrix $\rho_{1/2}$, along with transitions between the two subspaces mediated by transport processes. 
The current in a small static ($\sim$ millitesla) magnetic field responds sensitively to the relative orientation of the spin-polarized contact  and the divacancy ZFS axis. The MR  emerges from an induced spin polarization of the defect, a pathway to electrically initialize divacancy qubits. The MR signatures also enable sorting  divacancy ensembles into their fractional composition of each divacancy configuration.

% This approach avoids the restriction of low temperature required for spin-to-charge conversion in single quantum dots \cite{Elzerman2004}, and requires only modest static magnetic fields, providing an alternative path towards high temperature sensing in a broad range of environments. Leveraging spin-selective processes in this way can provide information on hyperfine coupling without the need for a FM contact in a variety of systems, such as double quantum dots\cite{Koppens2006} and defects at device interfaces\cite{Anders2020,Frantz2020}.

%%%%%%%%%%%% MODEL %%%%%%%%%%%%

The steady-state current is obtained from dynamics described by the stochastic Liouville equation,
 \begin{equation}
     \frac{d\rho(t)}{dt} = -\frac{i}{\hbar}[H,\rho(t)]-\mathcal{D}[\rho]+\mathcal{G}[\rho],
 \end{equation}
 where $\mathcal{D}[\rho]$ and $\mathcal{G}[\rho]$ represent the dissipators and generators of  Fock states that differ by one carrier (spin~1 or spin~1/2) respectively, with $d\rho/dt=0$. 
 The  spin Hamiltonians are, for the  spin~1/2  state,
 %singly occupied ionic divacancy, 
 \begin{equation}\label{eqn:IonicHamiltonian}
     H_\text{1/2} = g \mu_\text{B}\mathbf{B}\cdot\mathbf{s},
 \end{equation}
with $\mathbf{s}$   the  spin~1/2 operators and $\mathbf{B}$ the applied field, and for  the spin~1  state, %neutral divacancy
 \begin{equation}\label{eqn:NeutralHamiltonian}
     H_\text{1} = g \mu_\text{B}\mathbf{B}\cdot\mathbf{S} + D S_z^2-E(S_x^2-S_y^2),
 \end{equation}
with  $\mathbf{S} = \mathbf{s}_a+\mathbf{s}_b$   the  spin~1 operators (only one spin~1/2 is occupied in Eq.~(\ref{eqn:IonicHamiltonian}) so no distinguishing label is needed), and $D$ and $E$ are the longitudinal and transverse zero-field splitting parameters respectively. These  two Hamiltonians describe the coherent spin evolution of the density matrices in $\mathbf{B}$ and the ZFS terms.

$\gamma_\text{N}$ is the rate for orbitally incoherent hopping  from the nonmagnetic bulk to the defect.
The generator of the spin~1 state
 \begin{equation}
     \mathcal{G}_{1/2 \to 1} = \frac{2}{3}\gamma_\text{N} P_1(\mathbf{I}_{2\times 2} \otimes \rho_\text{1/2})P_1,
 \end{equation}
 where $P_1$ projects onto the spin-1 subspace and the factor of 2/3 normalizes the trace.  Generating a spin~1 state also dissipates the spin~1/2 state, according to 
 \begin{equation}
     \mathcal{D}_{1/2\to 1} = -\gamma_\text{N} \rho_{1/2}.
 \end{equation}
 
Dissipation of the neutral spin~1 state and  generation of the spin~1/2 ionized state occur via hopping with rate  $\gamma_\text{F}$ to the spin-selective FM contact with a magnetization operator $\hat{M}$.  The anti-commuting form of $\hat{M}$ correctly describes the decay of  coherence \cite{Haberkorn1976}:
 \begin{equation}
     \mathcal{D}_{1\to 1/2} =\gamma_\text{F}( \{\hat{M}^{(a)},\rho_1\}+\{\hat{M}^{(b)},\rho_1\}),
 \end{equation}
 where $\rho_1 = \rho_a\otimes\rho_b$ and
 \begin{equation}
     \hat{M}^{(j)} = \frac{1}{4}(\mathbf{I}_{4\times 4}+P\Sigma^{(j)}_x),
 \end{equation} 
 with $j=a,b$ corresponding to an individual spin-1/2 subspace, $P$ the in-plane polarization of the FM contact, and $\Sigma^{(j)}_i$ the $i^\text{th}$ 4x4 Pauli matrix for subspace $j$. For simplicity we  assume 100\%\ FM polarization  ($P=1$); lower polarization  decreases the MR contrast. The spin~1/2 manifold generation
 \begin{equation}
     \mathcal{G}_{1\to 1/2} = \gamma_\text{F}( \text{Tr}_a[\{\hat{M}^{(a)},\rho_1\}]+\text{Tr}_b [\{\hat{M}^{(b)},\rho_1\}]),
 \end{equation}
 where $\text{Tr}_j$ is the partial trace over the spin~1/2 subspace $j$. These terms lead to divacancy spin decoherence.
 
In addition to decoherence from transport, the spin can decohere within a single manifold through interactions with the local environment, characterized by the longitudinal relaxation time, $T_1$, and spin decoherence time $T_2$. For  $\rho_{1/2}$ the on-site dissipation is expressed with Lindblad terms:
\begin{equation}\label{eqn:Lindblad_spin1/2}
    \begin{aligned}
     L_1 = &\frac{1}{\sqrt{2 T_1}}
     \begin{pmatrix}
         0 & 1 \\
         0 & 0
     \end{pmatrix}\\
     L_2 = &\frac{1}{\sqrt{2 T_1}}
     \begin{pmatrix}
         0 & 0 \\
         1 & 0
     \end{pmatrix}\\
     L_3 = &\sqrt{\frac{1}{2 T_2}-\frac{1}{4 T_2}}
     \begin{pmatrix}
         1 & 0 \\
         0 & -1
     \end{pmatrix}.\\     
    \end{aligned}
\end{equation}
 Direct products between  two spin~1/2 sub-spaces extends these expressions to the spin~1 manifold with $i=1,2,3$.
\begin{equation}
    \begin{aligned}
     L^{(a)}_i = &P_1(L_i\otimes \mathbf{I}_{2\times 2})P_1\\
     L^{(b)}_i = &P_1(\mathbf{I}_{2\times 2}\otimes L_i)P_1,
    \end{aligned}
\end{equation}
% where $i=1,2,3$ and $L_i$ are given in Eq~(\ref{eqn:Lindblad_spin1/2}).
 Putting these expressions together yields:
 \begin{equation}\label{eqn:FullSLE}
     \begin{aligned}
      \frac{d\rho_{1/2}(t)}{dt} =&-\frac{i}{\hbar}[H_{1/2},\rho_{1/2}(t)]\\
      &- \mathcal{D}_{1/2\to 1}[\rho_{1/2}(t)]+ \mathcal{G}_{1\to 1/2}[\rho_1(t)]\\ 
      &+ \sum^3_{j=1} \bigg( L_j\rho_{1/2}(t)L^{\dagger}_j - \frac{1}{2}\{L^\dagger_jL_j,\rho_{1/2}(t)\}\bigg)\\
      \frac{d\rho_{1}(t)}{dt} =&-\frac{i}{\hbar}[H_{1},\rho_{1}(t)] - \mathcal{D}_{1\to 1/2}[\rho_{1}(t)] \\
      &+ \mathcal{G}_{1/2\to 1}[\rho_{1/2}(t)]\\
      &+ \sum^{3}_{j=1} \bigg( L^{(a)}_j\rho_{1}(t)L^{(a)\dagger}_j - \frac{1}{2}\{L^{(a)\dagger}_jL^{(a)}_j,\rho_{1}(t)\}\bigg)\\
      &+ \sum^{3}_{j=1} \bigg( L^{(b)}_j\rho_{1}(t)L^{(b)\dagger}_j - \frac{1}{2}\{L^{(b)\dagger}_jL^{(b)}_j,\rho_{1}(t)\}\bigg),\\
     \end{aligned}
 \end{equation}
 where $\{...\}$ represents  anti-commutation.

 The  operator corresponding to current onto the defect is proportional to $\mathcal{G}_{1/2 \to 1}$, and %corresponding to current 
 off of the defect is proportional to $\mathcal{D}_{1\to 1/2}$. They are
 \begin{equation}\label{eqn:CurrentOpOntoDefect}
     \hat{I}_\text{N} = e\mathcal{G}_{1/2 \to 1}=
     \frac{2e\gamma_\text{N}}{3} P_1(\mathbf{I}_{2\times 2} \otimes \rho_{1/2})P_1,
 \end{equation}
 \begin{equation}\label{eqn:CurrentOpOffOfDefect}
     \hat{I}_\text{F} =e\mathcal{D}_{1\to 1/2}= e\gamma_\text{F} ( \{\hat{M}^{(a)},\rho_1\}+\{\hat{M}^{(b)},\rho_1\}).
 \end{equation}
 The charge current is calculated from the trace of these operators, $I=\text{Tr}\hat{I}_\text{N}=\text{Tr}\hat{I}_\text{F}$.
 
%%%%%%%%%%%% RESULTS %%%%%%%%%%%%
We now apply this general formalism to $(hh)$ and $(kk)$ divacancies in 4H-SiC with the $c$~axis normal to the surface [Fig.~\ref{fig:SetupC3vMR}(a)]. These divacancies are $C_{3v}$ symmetric with defect axis along the c-axis of the crystal, and as a result the crystal field does not induce a transverse zero-field splitting. The applied bias also will not contribute to a transverse zero-field term, and thus $E=0$ in Eq.~(\ref{eqn:NeutralHamiltonian}).
As depicted in Fig.~\ref{fig:SetupC3vMR}(a), we consider $\mathbf{B} \parallel c$~axis and the  contact magnetization $\perp c$~axis.
In this high-symmetry configuration the current through an individual divacancy can be calculated analytically, and in the limit ($\gamma_\text{N}\to\infty$),  \begin{widetext}
 \begin{equation}
     I_{\gamma_\text{N}\to \infty}(\tilde{B})=e\gamma_\text{F}
     \bigg[\frac{1728(\tilde{B}^3-\tilde{B}\tilde{D}^2)^2+48(50\tilde{B}^4-3\tilde{B}^2\tilde{D}^2+9\tilde{D}^4)\gamma^2_\text{F} + 75(4\tilde{B}^2+3\tilde{D}^2)\gamma^4_\text{F}}
     {1728(\tilde{B}^3-\tilde{B}\tilde{D}^2)^2 + 16(166\tilde{B}^4+39\tilde{B}^2\tilde{D}^2+27\tilde{D}^4)\gamma^2_\text{F}+(700\tilde{B}^2+417\tilde{D}^2)\gamma^4_\text{F}+100\gamma^6_\text{F}}\bigg],
 \end{equation}
 \end{widetext}
 where $\tilde{B}=g\mu_\text{B}B/\hbar$ and $\tilde{D}=D/\hbar$.

\begin{figure}[htbp]
\begin{centering}
       \includegraphics[scale=0.4]{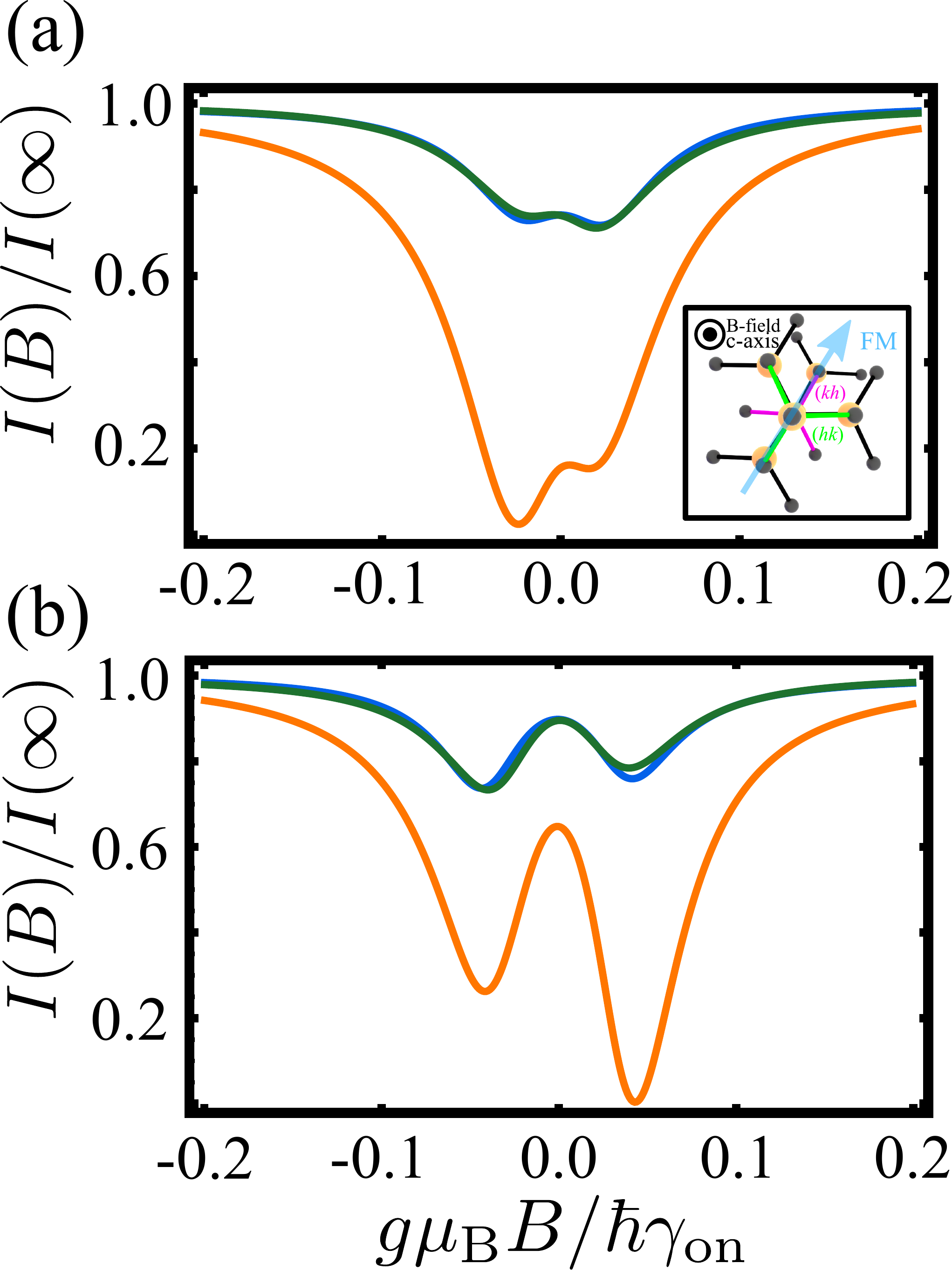}
       \caption[width=\textwidth]
{(a) MR for  \textit{(kh)} divacancies with  azimuthal angle of 0 (orange), $2\pi/3$ (blue), and $4\pi/3$ (green) relative to the FM  magnetization. (b) MR for  \textit{(hk)} divacancy with  azimuthal angle of $\pi/3$ (blue), $\pi$ (orange), and $5\pi/3$ (green). For both $\gamma_\text{F}/\gamma_\text{N}=0.02$. Inset: sketch of \textit{(kh)} and \textit{(hk)} divacancy orientation and the FM magnetization viewed along the $c$-axis.}\label{fig:C1hMR}
       \end{centering}
\end{figure}

\begin{figure*}[htbp!]\label{fig:CompositCurrent}
\begin{centering}
       \includegraphics[width=\textwidth,scale=0.5]{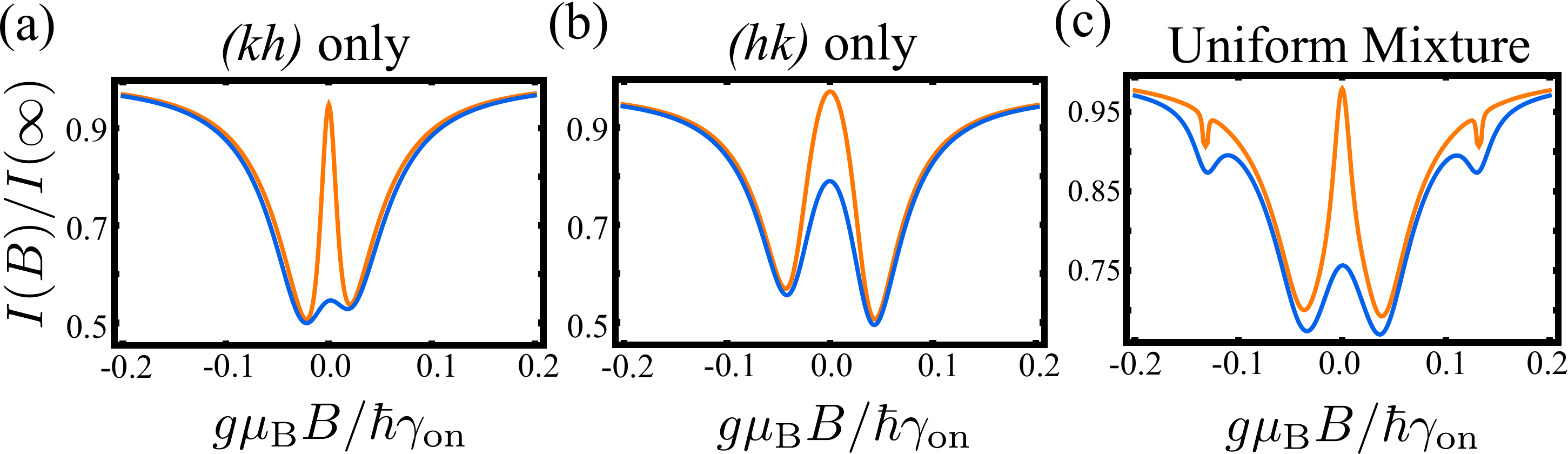}
       \caption[width=\textwidth]
{Ensemble magnetoresistance for a uniform distribution of \textit{(kh)} (a), \textit{(hk)} (b), and each of the four (c) divacancies for two different hopping ratios: $\gamma_\text{F}/\gamma_\text{N} = 0.02$ (blue) and $\gamma_\text{F}/\gamma_\text{N} = 0.002$ (orange). }\label{fig:EnsembleMR}
       \end{centering}
\end{figure*}

Current from the bulk 4H-SiC substrate through the defect will respond to an applied magnetic field due to the non-equilibrium spin-spin correlation between the magnetization of the planar contact and the spin state of the defect. {The finite-field features in Fig.~\ref{fig:SetupC3vMR}(d) reflect the induced polarization of the defect spin as a result of these non-equilibrium correlations. }

In the Zeeman basis parallel to the defect axis and $\mathbf{B} = 0$ the ZFS  splits the $m_s=0$ from the $m_s=\pm 1$ states by  $D$ [Fig.~\ref{fig:SetupC3vMR}(c)]. For $\mathbf{B}\ne 0$, the $m_s=\pm 1$ states split further from the Zeeman effect. When $g\mu_\text{B}|B|=|D|$, a degeneracy occurs between the $m_s=0$ and a $|m_s|=1$. There is a preferential conduction of spin parallel to the quantization axis of the FM, and in the ideal case ($P=1$) \textit{only} spins that are parallel to the quantization axis can hop. This leads to dynamical polarization of the divacancy spin, with spin orientation least likely to hop (anti-parallel to the FM). The dynamical polarization manifests as a current dip, caused by degenerate eigenstates that can coherently sum to produce a ``bottlenecked'' state [Fig.~\ref{fig:SetupC3vMR}(d)]. This bottleneck state is $(2/3)^{1/2}|T_+\rangle -(1/3)^{1/2}|T_{\mathrm 0}\rangle$ {\it in the spin basis of the FM contact}.% \textcolor{red}{an aside, but do you know what the bottleneck state is at resonance?}

The azimuthal magnetic orientation of the planar contact is irrelevant for the magnetoresistance of the \textit{(hh)} and \textit{(kk)} divacancies in this configuration. For \textit{(hk)} and \textit{(kh)} orientations shown in Fig.~\ref{fig:SetupC3vMR}(b), however, the defect axes are oblique to the $c$~axis. For the SiC crystal only three azimuthal orientations are possible for each of the two basal divacancies. These are shown in Fig.~\ref{fig:C1hMR}(a). These defects have $C_{1h}$ symmetry and the crystal field \textit{does} provide transverse zero-field splitting ($E\neq 0$).

\begin{figure}[htbp!]
\begin{centering}
       \includegraphics[width=\columnwidth,scale=0.5]{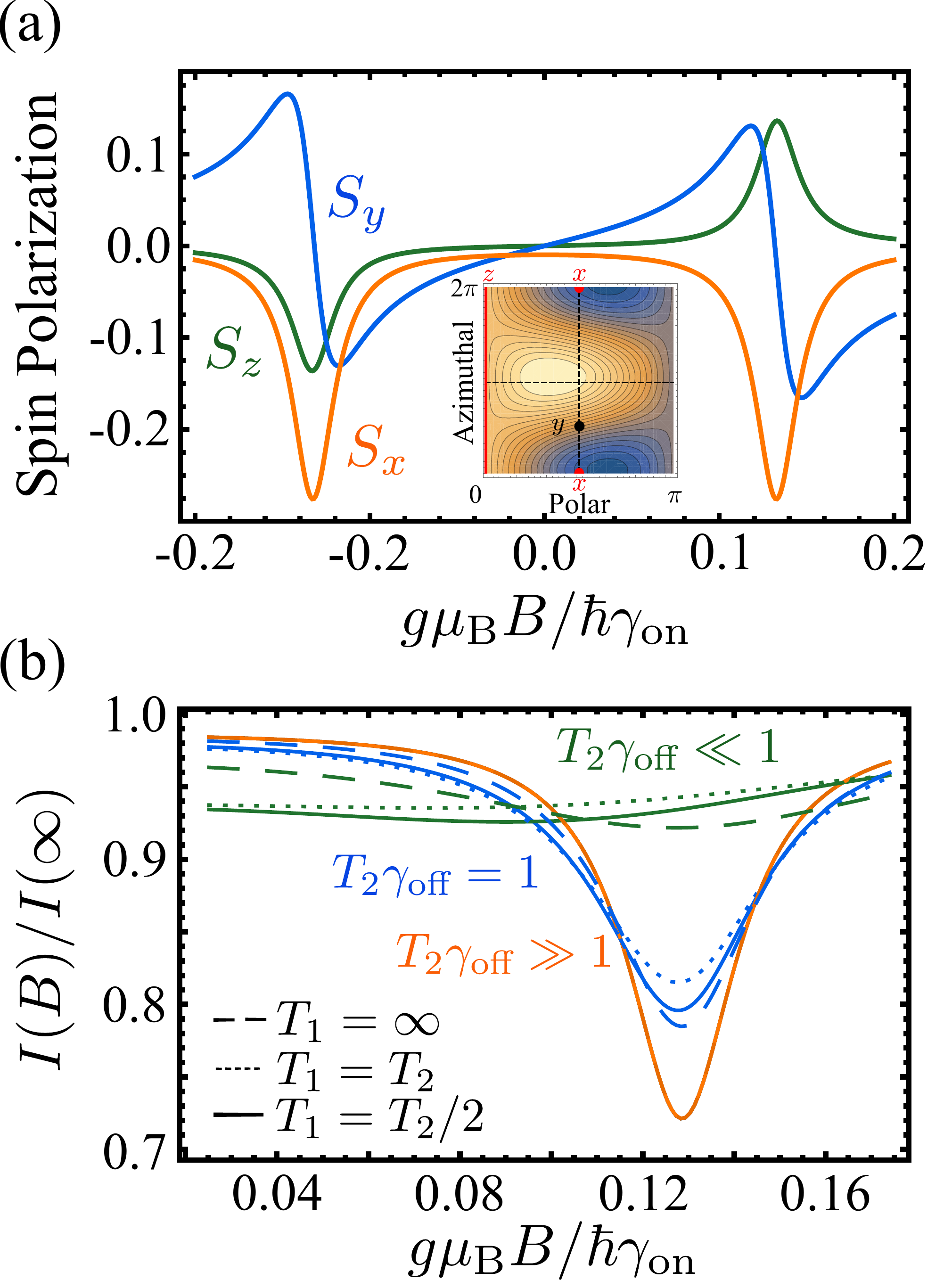}
       \caption[width=\textwidth]
{(a) Spin polarization of  \textit{(hh)} divacancy.
(Inset) contour plot of  $B>0$ total spin polarization, with light color indicating a maximum. Points of interest are labeled in the defect orientation basis. $\hat z \parallel$  defect axis and  $\perp$ to the  interface, with  FM polarization $\parallel \hat x$. The maximum polarization is $0.3$ along the $(\theta,\phi)=(1.11,3.26)$ direction. (b) Positive field current dip of  \textit{(hh)} divacancy with negligible (orange, $T_2\gamma_\text{N}=10^{5}$), considerable (blue, $T_2\gamma_\text{N}=100$), and substantial (green, $T_2\gamma_\text{N}=10$) on-site decoherence, assuming three different $T_1$'s. Decoherence broadens the signal whereas finite $T_1$ shifts to smaller $B$.  $\gamma_\text{F}/\gamma_\text{N}=0.02$. }\label{fig:SpinPolAndOnSiteDeco}
       \end{centering}
\end{figure}

Figure ~\ref{fig:C1hMR}(bc) shows the steady-state current through an individual \textit{(kh)} and \textit{(hk)} divacancy respectively. The orientation of the FM is fixed in-plane with the same azimuthal angle as one of the \textit{(kh)} divacancies, sketched as the blue arrow in Fig.~\ref{fig:C1hMR}(a). We assume a longitudinal zero-field splitting of $D/\hbar\gamma_\text{N} = 0.13$ and a transverse zero-field splitting of $E/\hbar\gamma_\text{N}= 0.0018$ for the \textit{(kh)}, and $D/\hbar\gamma_\text{N} = 0.12$ and $E/\hbar\gamma_\text{N}= 0.0082$ for the \textit{(hk)}. The overlaid traces in Fig.~\ref{fig:C1hMR}(bc) show the variation in MR for each of the three orientations. For both \textit{(kh)} and \textit{(hk)} the largest polarization occurs for the defect axis azimuthally parallel or anti-parallel to the FM polarization (orange). It can also be seen that the larger asymmetric feature depends on the  angle between the defect and  FM axes. The Hamiltonian in Eq.~(\ref{eqn:NeutralHamiltonian}) is independent of the FM orientation, and thus the dip condition is independent of the relative orientation of the defect axis. This is reflected in the MR features for defects in identical crystal environments. 
Conversely, differences in the zero-field splitting parameters provide a means of identifying the local crystal environment through measurement of the finite-field current dips. The scale of the defect spin polarization and the degree of asymmetry depend on the relative orientation of the defect axis, providing a distinct signature for each unique orientation of a divacancy. 

A mixed ensemble of non-interacting divacancies  exhibits MR with increased complexity due to each distinct contributing signature. 
Interacting defects are not considered here, but the example of two spin-1/2 centers interacting through exchange has been shown to modify the magnetoresistance\cite{McMillan2020}. The results described here describe ensembles with typical defect spacings $\sim 10$~nm. Dipolar interactions at this distance are of the order of 500~kHz and exchange interactions even smaller --- unresolvable for the currents considered here.
The MR thus directly provides the concentration of contributing defects of a given orientation. The total current $I(B)_\text{comp} = \sum_i w_i I_i(B)$, where $w_i$ indicates the fractional population and $i$ labels divacancy configuration. 

Figure~\ref{fig:EnsembleMR} shows the composite signal for \textit{(kh)}:  ($w_{(kh),0}=w_{(kh),2\pi/3}=w_{(kh),4\pi/3}=1/3$), \textit{(hk)}: ($w_{(hk),\pi/3}=w_{(hk),\pi}=w_{(hk),5\pi/3}=1/3$), and a uniform mixture of the four types of divacancies ($w_{i}=1/8$) respectively. The two traces represent hopping ratios that differ by an order of magnitude. For  $\gamma_\text{N}\approx\text{ns}^{-1}$, the applied magnetic field required to resolve the zero-field energies is $\sim$~mT.  
The smallest energy scale of the system is set by  $\gamma_\text{F}$; in Fig.~\ref{fig:EnsembleMR} the higher resolution is obtained when  $\gamma_\text{F}\sim 2$~MHz, corresponding to single-defect currents of 0.3~pA and $T_2> 10 \mu$s. The total current  scales with the defect number, and such $T_2$'s are typical for divacancies in 4H-SiC \cite{Christle2015}. 

Divacancy spin decoherence further influences the features discussed above. Figure~\ref{fig:SpinPolAndOnSiteDeco}(a) shows the spin  for an \textit{(hh)} divacancy. The FM polarization $\parallel \hat x$, $\mathbf{B}\parallel \hat z$ (the defect axis). At the current dip  the spin polarization is primarily in the $xz$ plane, indicating an axis of maximum polarization oblique to the defect axis. The inset of Fig.~\ref{fig:SpinPolAndOnSiteDeco}(a) is a contour plot of the total spin polarization at the positive $B$ field current dip in the defect basis. The  $+\hat x$ and $+\hat z$ directions are red and the $+\hat y$  is black. Lighter colors indicate positive values and darker colors indicate negative values. The axis of maximum polarization has an orientation of $(\theta,\phi)=(1.11,3.26)$ radians in the $xz$ basis with a magnitude of $|S_\text{max}|=0.3$.

Fig.~\ref{fig:SpinPolAndOnSiteDeco}(b) shows the steady-state current dip at $+B$ for a single $(hh)$ defect from Eq.~(\ref{eqn:FullSLE})'s solution.
We consider three different regimes of operation: weak ($T_2\gamma_\text{N}=10^5$, orange), moderate ($T_2\gamma_\text{N}=10^2$, blue), and strong ($T_2\gamma_\text{N}=10$, green) dephasing, with a fixed hopping ratio $\gamma_\text{F}/\gamma_\text{N} = 0.02$. In each regime we consider three relationships between the longitudinal relaxation $T_1$ and the decoherence time $T_2$ indicated by the line type. In the weak regime, all three  relations are equivalent and the feature reduces to the trace for negligible on-site $T_2$. The moderate and strong regimes show a general broadening of the  feature, with $T_1$ shifting the feature to lower field, as is most apparent in the strong regime.

\begin{acknowledgments}
This material is based on work supported by the U. S. Department of Energy, Office of Science, Office of Basic Energy Sciences, under Award Number DE-SC0016379. We would like to acknowledge N. J. Harmon and D. R. Candido for useful and valuable discussions.
\end{acknowledgments}

\bibliography{central-refs.bib}

\begin{thebibliography}{32}%
\makeatletter
\providecommand \@ifxundefined [1]{%
 \@ifx{#1\undefined}
}%
\providecommand \@ifnum [1]{%
 \ifnum #1\expandafter \@firstoftwo
 \else \expandafter \@secondoftwo
 \fi
}%
\providecommand \@ifx [1]{%
 \ifx #1\expandafter \@firstoftwo
 \else \expandafter \@secondoftwo
 \fi
}%
\providecommand \natexlab [1]{#1}%
\providecommand \enquote  [1]{``#1''}%
\providecommand \bibnamefont  [1]{#1}%
\providecommand \bibfnamefont [1]{#1}%
\providecommand \citenamefont [1]{#1}%
\providecommand \href@noop [0]{\@secondoftwo}%
\providecommand \href [0]{\begingroup \@sanitize@url \@href}%
\providecommand \@href[1]{\@@startlink{#1}\@@href}%
\providecommand \@@href[1]{\endgroup#1\@@endlink}%
\providecommand \@sanitize@url [0]{\catcode `\\12\catcode `\$12\catcode
  `\&12\catcode `\#12\catcode `\^12\catcode `\_12\catcode `\%12\relax}%
\providecommand \@@startlink[1]{}%
\providecommand \@@endlink[0]{}%
\providecommand \url  [0]{\begingroup\@sanitize@url \@url }%
\providecommand \@url [1]{\endgroup\@href {#1}{\urlprefix }}%
\providecommand \urlprefix  [0]{URL }%
\providecommand \Eprint [0]{\href }%
\providecommand \doibase [0]{http://dx.doi.org/}%
\providecommand \selectlanguage [0]{\@gobble}%
\providecommand \bibinfo  [0]{\@secondoftwo}%
\providecommand \bibfield  [0]{\@secondoftwo}%
\providecommand \translation [1]{[#1]}%
\providecommand \BibitemOpen [0]{}%
\providecommand \bibitemStop [0]{}%
\providecommand \bibitemNoStop [0]{.\EOS\space}%
\providecommand \EOS [0]{\spacefactor3000\relax}%
\providecommand \BibitemShut  [1]{\csname bibitem#1\endcsname}%
\let\auto@bib@innerbib\@empty
%</preamble>
\bibitem [{\citenamefont {Meier}\ and\ \citenamefont
  {Zachachrenya}(1984)}]{Meier1984}%
  \BibitemOpen
  \bibfield  {author} {\bibinfo {author} {\bibfnamefont {F.}~\bibnamefont
  {Meier}}\ and\ \bibinfo {author} {\bibfnamefont {B.~P.}\ \bibnamefont
  {Zachachrenya}},\ }\href@noop {} {\emph {\bibinfo {title} {Optical
  Orientation: Modern Problems in Condensed Matter Science}}},\ Vol.~\bibinfo
  {volume} {8}\ (\bibinfo  {publisher} {North-Holland},\ \bibinfo {address}
  {Amsterdam},\ \bibinfo {year} {1984})\BibitemShut {NoStop}%
\bibitem [{\citenamefont {Awschalom}\ \emph {et~al.}(2002)\citenamefont
  {Awschalom}, \citenamefont {Samarth},\ and\ \citenamefont
  {Loss}}]{Awschalom2002}%
  \BibitemOpen
  \bibinfo {editor} {\bibfnamefont {D.~D.}\ \bibnamefont {Awschalom}}, \bibinfo
  {editor} {\bibfnamefont {N.}~\bibnamefont {Samarth}}, \ and\ \bibinfo
  {editor} {\bibfnamefont {D.}~\bibnamefont {Loss}},\ eds.,\ \href@noop {}
  {\emph {\bibinfo {title} {Semiconductor Spintronics and Quantum
  Computation}}}\ (\bibinfo  {publisher} {Springer Verlag},\ \bibinfo {address}
  {Heidelberg},\ \bibinfo {year} {2002})\BibitemShut {NoStop}%
\bibitem [{\citenamefont {Hanson}\ \emph {et~al.}(2007)\citenamefont {Hanson},
  \citenamefont {Kouwenhoven}, \citenamefont {Petta}, \citenamefont {Tarucha},\
  and\ \citenamefont {Vandersypen}}]{Hanson2007}%
  \BibitemOpen
  \bibfield  {author} {\bibinfo {author} {\bibfnamefont {R.}~\bibnamefont
  {Hanson}}, \bibinfo {author} {\bibfnamefont {L.~P.}\ \bibnamefont
  {Kouwenhoven}}, \bibinfo {author} {\bibfnamefont {J.~R.}\ \bibnamefont
  {Petta}}, \bibinfo {author} {\bibfnamefont {S.}~\bibnamefont {Tarucha}}, \
  and\ \bibinfo {author} {\bibfnamefont {L.~M.~K.}\ \bibnamefont
  {Vandersypen}},\ }\bibfield  {title} {\enquote {\bibinfo {title} {Spins in
  few-electron quantum dots},}\ }\href@noop {} {\bibfield  {journal} {\bibinfo
  {journal} {Rev. Mod. Phys.}\ }\textbf {\bibinfo {volume} {79}},\ \bibinfo
  {pages} {1217} (\bibinfo {year} {2007})}\BibitemShut {NoStop}%
\bibitem [{\citenamefont {Loss}\ and\ \citenamefont
  {DiVincenzo}(1998)}]{Loss1998}%
  \BibitemOpen
  \bibfield  {author} {\bibinfo {author} {\bibfnamefont {D.}~\bibnamefont
  {Loss}}\ and\ \bibinfo {author} {\bibfnamefont {D.~P.}\ \bibnamefont
  {DiVincenzo}},\ }\bibfield  {title} {\enquote {\bibinfo {title} {Quantum
  computation with quantum dots},}\ }\href@noop {} {\bibfield  {journal}
  {\bibinfo  {journal} {\pra}\ }\textbf {\bibinfo {volume} {57}},\ \bibinfo
  {pages} {120--126} (\bibinfo {year} {1998})}\BibitemShut {NoStop}%
\bibitem [{\citenamefont {Kane}(1998)}]{Kane1998}%
  \BibitemOpen
  \bibfield  {author} {\bibinfo {author} {\bibfnamefont {B.~E.}\ \bibnamefont
  {Kane}},\ }\bibfield  {title} {\enquote {\bibinfo {title} {A silicon-based
  nuclear spin quantum computer},}\ }\href@noop {} {\bibfield  {journal}
  {\bibinfo  {journal} {Nature}\ }\textbf {\bibinfo {volume} {393}},\ \bibinfo
  {pages} {133--137} (\bibinfo {year} {1998})}\BibitemShut {NoStop}%
\bibitem [{\citenamefont {Balasubramanian}\ \emph {et~al.}(2009)\citenamefont
  {Balasubramanian}, \citenamefont {Neumann}, \citenamefont {D.~Twitchen},
  \citenamefont {Markham}, \citenamefont {Kolesov}, \citenamefont {Mizuochi},
  \citenamefont {Isoya}, \citenamefont {Achard}, \citenamefont {Beck},
  \citenamefont {Tissler}, \citenamefont {Jacques}, \citenamefont {Hemmer},
  \citenamefont {Jelezko},\ and\ \citenamefont
  {Wrachtrup}}]{Balasubramanian2009}%
  \BibitemOpen
  \bibfield  {author} {\bibinfo {author} {\bibfnamefont {G.}~\bibnamefont
  {Balasubramanian}}, \bibinfo {author} {\bibfnamefont {P.}~\bibnamefont
  {Neumann}}, \bibinfo {author} {\bibfnamefont {D.}~\bibnamefont
  {D.~Twitchen}}, \bibinfo {author} {\bibfnamefont {M.}~\bibnamefont
  {Markham}}, \bibinfo {author} {\bibfnamefont {R.}~\bibnamefont {Kolesov}},
  \bibinfo {author} {\bibfnamefont {N.}~\bibnamefont {Mizuochi}}, \bibinfo
  {author} {\bibfnamefont {J.}~\bibnamefont {Isoya}}, \bibinfo {author}
  {\bibfnamefont {J.}~\bibnamefont {Achard}}, \bibinfo {author} {\bibfnamefont
  {J.}~\bibnamefont {Beck}}, \bibinfo {author} {\bibfnamefont {J.}~\bibnamefont
  {Tissler}}, \bibinfo {author} {\bibfnamefont {V.}~\bibnamefont {Jacques}},
  \bibinfo {author} {\bibfnamefont {P.~R.}\ \bibnamefont {Hemmer}}, \bibinfo
  {author} {\bibfnamefont {F.}~\bibnamefont {Jelezko}}, \ and\ \bibinfo
  {author} {\bibfnamefont {J.}~\bibnamefont {Wrachtrup}},\ }\bibfield  {title}
  {\enquote {\bibinfo {title} {Ultralong spin coherence time in isotopically
  engineered diamond},}\ }\href@noop {} {\bibfield  {journal} {\bibinfo
  {journal} {Nat. Mater.}\ }\textbf {\bibinfo {volume} {8}},\ \bibinfo {pages}
  {383} (\bibinfo {year} {2009})}\BibitemShut {NoStop}%
\bibitem [{\citenamefont {Koenraad}\ and\ \citenamefont
  {Flatt{\'e}}(2011)}]{Koenraad2011}%
  \BibitemOpen
  \bibfield  {author} {\bibinfo {author} {\bibfnamefont {P.~M.}\ \bibnamefont
  {Koenraad}}\ and\ \bibinfo {author} {\bibfnamefont {M.~E.}\ \bibnamefont
  {Flatt{\'e}}},\ }\bibfield  {title} {\enquote {\bibinfo {title} {Single
  dopants in semiconductors},}\ }\href@noop {} {\bibfield  {journal} {\bibinfo
  {journal} {Nat. Mater.}\ }\textbf {\bibinfo {volume} {10}},\ \bibinfo {pages}
  {91} (\bibinfo {year} {2011})}\BibitemShut {NoStop}%
\bibitem [{\citenamefont {Degen}\ \emph {et~al.}(2017)\citenamefont {Degen},
  \citenamefont {Reinhard},\ and\ \citenamefont {Cappellaro}}]{Degen2017}%
  \BibitemOpen
  \bibfield  {author} {\bibinfo {author} {\bibfnamefont {C.~L.}\ \bibnamefont
  {Degen}}, \bibinfo {author} {\bibfnamefont {F.}~\bibnamefont {Reinhard}}, \
  and\ \bibinfo {author} {\bibfnamefont {P.}~\bibnamefont {Cappellaro}},\
  }\bibfield  {title} {\enquote {\bibinfo {title} {Quantum sensing},}\
  }\href@noop {} {\bibfield  {journal} {\bibinfo  {journal} {Rev. Mod. Phys.}\
  }\textbf {\bibinfo {volume} {89}},\ \bibinfo {pages} {035002} (\bibinfo
  {year} {2017})}\BibitemShut {NoStop}%
\bibitem [{\citenamefont {Nielsen}\ and\ \citenamefont
  {Chuang}(2010)}]{Nielsen2010}%
  \BibitemOpen
  \bibfield  {author} {\bibinfo {author} {\bibfnamefont {M.}~\bibnamefont
  {Nielsen}}\ and\ \bibinfo {author} {\bibfnamefont {I.}~\bibnamefont
  {Chuang}},\ }\href@noop {} {\emph {\bibinfo {title} {Quantum Computation and
  Quantum Information}}}\ (\bibinfo  {publisher} {Cambridge University Press},\
  \bibinfo {address} {New York},\ \bibinfo {year} {2010})\BibitemShut {NoStop}%
\bibitem [{\citenamefont {Aharonovich}\ \emph {et~al.}(2016)\citenamefont
  {Aharonovich}, \citenamefont {Englund},\ and\ \citenamefont
  {Toth}}]{Aharonovich2016}%
  \BibitemOpen
  \bibfield  {author} {\bibinfo {author} {\bibfnamefont {I.}~\bibnamefont
  {Aharonovich}}, \bibinfo {author} {\bibfnamefont {D.}~\bibnamefont
  {Englund}}, \ and\ \bibinfo {author} {\bibfnamefont {M.}~\bibnamefont
  {Toth}},\ }\bibfield  {title} {\enquote {\bibinfo {title} {Solid-state
  single-photon emitters},}\ }\href@noop {} {\bibfield  {journal} {\bibinfo
  {journal} {Nat. Photonics}\ }\textbf {\bibinfo {volume} {10}},\ \bibinfo
  {pages} {631--641} (\bibinfo {year} {2016})}\BibitemShut {NoStop}%
\bibitem [{\citenamefont {Schofield}\ \emph {et~al.}(2003)\citenamefont
  {Schofield}, \citenamefont {Curson}, \citenamefont {Simmons}, \citenamefont
  {Rue\ss{}}, \citenamefont {Hallam}, \citenamefont {Oberbeck},\ and\
  \citenamefont {Clark}}]{Schofield2003}%
  \BibitemOpen
  \bibfield  {author} {\bibinfo {author} {\bibfnamefont {S.~R.}\ \bibnamefont
  {Schofield}}, \bibinfo {author} {\bibfnamefont {N.~J.}\ \bibnamefont
  {Curson}}, \bibinfo {author} {\bibfnamefont {M.~Y.}\ \bibnamefont {Simmons}},
  \bibinfo {author} {\bibfnamefont {F.~J.}\ \bibnamefont {Rue\ss{}}}, \bibinfo
  {author} {\bibfnamefont {T.}~\bibnamefont {Hallam}}, \bibinfo {author}
  {\bibfnamefont {L.}~\bibnamefont {Oberbeck}}, \ and\ \bibinfo {author}
  {\bibfnamefont {R.~G.}\ \bibnamefont {Clark}},\ }\bibfield  {title} {\enquote
  {\bibinfo {title} {Atomically precise placement of single dopants in
  \hbox{Si}},}\ }\href@noop {} {\bibfield  {journal} {\bibinfo  {journal}
  {Phys. Rev. Lett.}\ }\textbf {\bibinfo {volume} {91}},\ \bibinfo {pages}
  {136104} (\bibinfo {year} {2003})}\BibitemShut {NoStop}%
\bibitem [{\citenamefont {Fuechsle}\ \emph {et~al.}(2012)\citenamefont
  {Fuechsle}, \citenamefont {Miwa}, \citenamefont {Mahapatra}, \citenamefont
  {Ryu}, \citenamefont {Lee}, \citenamefont {Warschkow}, \citenamefont
  {Hollenberg}, \citenamefont {Klimeck},\ and\ \citenamefont
  {Simmons}}]{Fuechsle2012}%
  \BibitemOpen
  \bibfield  {author} {\bibinfo {author} {\bibfnamefont {M.}~\bibnamefont
  {Fuechsle}}, \bibinfo {author} {\bibfnamefont {J.~A.}\ \bibnamefont {Miwa}},
  \bibinfo {author} {\bibfnamefont {S.}~\bibnamefont {Mahapatra}}, \bibinfo
  {author} {\bibfnamefont {H.}~\bibnamefont {Ryu}}, \bibinfo {author}
  {\bibfnamefont {S.}~\bibnamefont {Lee}}, \bibinfo {author} {\bibfnamefont
  {O.}~\bibnamefont {Warschkow}}, \bibinfo {author} {\bibfnamefont {L.~C.~L.}\
  \bibnamefont {Hollenberg}}, \bibinfo {author} {\bibfnamefont
  {G.}~\bibnamefont {Klimeck}}, \ and\ \bibinfo {author} {\bibfnamefont
  {M.~Y.}\ \bibnamefont {Simmons}},\ }\bibfield  {title} {\enquote {\bibinfo
  {title} {A single-atom transistor},}\ }\href@noop {} {\bibfield  {journal}
  {\bibinfo  {journal} {Nat. Nanotechnol.}\ }\textbf {\bibinfo {volume} {7}},\
  \bibinfo {pages} {242--246} (\bibinfo {year} {2012})}\BibitemShut {NoStop}%
\bibitem [{\citenamefont {Schofield}\ \emph {et~al.}(2013)\citenamefont
  {Schofield}, \citenamefont {Studer}, \citenamefont {Hirjibehedin},
  \citenamefont {Curson}, \citenamefont {Aeppli},\ and\ \citenamefont
  {Bowler}}]{Schofield2013}%
  \BibitemOpen
  \bibfield  {author} {\bibinfo {author} {\bibfnamefont {S.~R.}\ \bibnamefont
  {Schofield}}, \bibinfo {author} {\bibfnamefont {P.}~\bibnamefont {Studer}},
  \bibinfo {author} {\bibfnamefont {C.~F.}\ \bibnamefont {Hirjibehedin}},
  \bibinfo {author} {\bibfnamefont {N.~J.}\ \bibnamefont {Curson}}, \bibinfo
  {author} {\bibfnamefont {G.}~\bibnamefont {Aeppli}}, \ and\ \bibinfo {author}
  {\bibfnamefont {D.~R.}\ \bibnamefont {Bowler}},\ }\bibfield  {title}
  {\enquote {\bibinfo {title} {Quantum engineering at the silicon surface using
  dangling bonds},}\ }\href@noop {} {\bibfield  {journal} {\bibinfo  {journal}
  {Nat. Commun.}\ }\textbf {\bibinfo {volume} {4}},\ \bibinfo {pages} {1649}
  (\bibinfo {year} {2013})}\BibitemShut {NoStop}%
\bibitem [{\citenamefont {Lee}\ and\ \citenamefont {Gupta}(2019)}]{Lee2019}%
  \BibitemOpen
  \bibfield  {author} {\bibinfo {author} {\bibfnamefont {D.}~\bibnamefont
  {Lee}}\ and\ \bibinfo {author} {\bibfnamefont {J.~A.}\ \bibnamefont
  {Gupta}},\ }\bibfield  {title} {\enquote {\bibinfo {title} {Perspectives on
  deterministic control of quantum point defects by scanned probes},}\
  }\href@noop {} {\bibfield  {journal} {\bibinfo  {journal} {Nanophotonics}\
  }\textbf {\bibinfo {volume} {8}},\ \bibinfo {pages} {2033} (\bibinfo {year}
  {2019})}\BibitemShut {NoStop}%
\bibitem [{\citenamefont {Achal}\ \emph {et~al.}(2018)\citenamefont {Achal},
  \citenamefont {Rashidi}, \citenamefont {Croshaw}, \citenamefont {Churchill},
  \citenamefont {Taucer}, \citenamefont {Huff}, \citenamefont {Cloutier},
  \citenamefont {Pitters},\ and\ \citenamefont {Wolkow}}]{Achal2018}%
  \BibitemOpen
  \bibfield  {author} {\bibinfo {author} {\bibfnamefont {R.}~\bibnamefont
  {Achal}}, \bibinfo {author} {\bibfnamefont {M.}~\bibnamefont {Rashidi}},
  \bibinfo {author} {\bibfnamefont {J.}~\bibnamefont {Croshaw}}, \bibinfo
  {author} {\bibfnamefont {D.}~\bibnamefont {Churchill}}, \bibinfo {author}
  {\bibfnamefont {M.}~\bibnamefont {Taucer}}, \bibinfo {author} {\bibfnamefont
  {T.}~\bibnamefont {Huff}}, \bibinfo {author} {\bibfnamefont {M.}~\bibnamefont
  {Cloutier}}, \bibinfo {author} {\bibfnamefont {J.}~\bibnamefont {Pitters}}, \
  and\ \bibinfo {author} {\bibfnamefont {R.~A.}\ \bibnamefont {Wolkow}},\
  }\bibfield  {title} {\enquote {\bibinfo {title} {Lithography for robust and
  editable atomic-scale silicon devices and memories},}\ }\href@noop {}
  {\bibfield  {journal} {\bibinfo  {journal} {Nat. Commun.}\ }\textbf {\bibinfo
  {volume} {9}},\ \bibinfo {pages} {2778} (\bibinfo {year} {2018})}\BibitemShut
  {NoStop}%
\bibitem [{\citenamefont {Bradac}\ \emph {et~al.}(2019)\citenamefont {Bradac},
  \citenamefont {Gao}, \citenamefont {Forneris}, \citenamefont {Trusheim},\
  and\ \citenamefont {Aharonovich}}]{Bradac2019}%
  \BibitemOpen
  \bibfield  {author} {\bibinfo {author} {\bibfnamefont {C.}~\bibnamefont
  {Bradac}}, \bibinfo {author} {\bibfnamefont {W.}~\bibnamefont {Gao}},
  \bibinfo {author} {\bibfnamefont {J.}~\bibnamefont {Forneris}}, \bibinfo
  {author} {\bibfnamefont {M.~E.}\ \bibnamefont {Trusheim}}, \ and\ \bibinfo
  {author} {\bibfnamefont {I.}~\bibnamefont {Aharonovich}},\ }\bibfield
  {title} {\enquote {\bibinfo {title} {Quantum nanophotonics with group {IV}
  defects in diamond},}\ }\href@noop {} {\bibfield  {journal} {\bibinfo
  {journal} {Nat. Commun.}\ }\textbf {\bibinfo {volume} {10}},\ \bibinfo
  {pages} {5625} (\bibinfo {year} {2019})}\BibitemShut {NoStop}%
\bibitem [{\citenamefont {Stock}\ \emph {et~al.}(2020)\citenamefont {Stock},
  \citenamefont {Warschkow}, \citenamefont {Constantinou}, \citenamefont {Li},
  \citenamefont {Fearn}, \citenamefont {Crane}, \citenamefont {Hofmann},
  \citenamefont {K\"olker}, \citenamefont {McKenzie}, \citenamefont
  {Schofield},\ and\ \citenamefont {Curson}}]{Stock2020}%
  \BibitemOpen
  \bibfield  {author} {\bibinfo {author} {\bibfnamefont {Taylor J.~Z.}\
  \bibnamefont {Stock}}, \bibinfo {author} {\bibfnamefont {Oliver}\
  \bibnamefont {Warschkow}}, \bibinfo {author} {\bibfnamefont {Procopios~C.}\
  \bibnamefont {Constantinou}}, \bibinfo {author} {\bibfnamefont {Juerong}\
  \bibnamefont {Li}}, \bibinfo {author} {\bibfnamefont {Sarah}\ \bibnamefont
  {Fearn}}, \bibinfo {author} {\bibfnamefont {Eleanor}\ \bibnamefont {Crane}},
  \bibinfo {author} {\bibfnamefont {Emily V.~S.}\ \bibnamefont {Hofmann}},
  \bibinfo {author} {\bibfnamefont {Alexander}\ \bibnamefont {K\"olker}},
  \bibinfo {author} {\bibfnamefont {David~R.}\ \bibnamefont {McKenzie}},
  \bibinfo {author} {\bibfnamefont {Steven~R.}\ \bibnamefont {Schofield}}, \
  and\ \bibinfo {author} {\bibfnamefont {Neil~J.}\ \bibnamefont {Curson}},\
  }\bibfield  {title} {\enquote {\bibinfo {title} {Atomic-scale patterning of
  arsenic in silicon by scanning tunneling microscopy},}\ }\href {\doibase
  10.1021/acsnano.9b08943} {\bibfield  {journal} {\bibinfo  {journal} {ACS
  Nano}\ }\textbf {\bibinfo {volume} {14}},\ \bibinfo {pages} {3316--3327}
  (\bibinfo {year} {2020})},\ \bibinfo {note} {pMID: 32142256}\BibitemShut
  {NoStop}%
\bibitem [{\citenamefont {Baranov}\ \emph {et~al.}(2005)\citenamefont
  {Baranov}, \citenamefont {Il’in}, \citenamefont {Mokhov}, \citenamefont
  {Muzafarova}, \citenamefont {Orlinskii},\ and\ \citenamefont
  {Schmidt}}]{Baranov2005}%
  \BibitemOpen
  \bibfield  {author} {\bibinfo {author} {\bibfnamefont {P.~G.}\ \bibnamefont
  {Baranov}}, \bibinfo {author} {\bibfnamefont {I.~V.}\ \bibnamefont
  {Il’in}}, \bibinfo {author} {\bibfnamefont {E.~N.}\ \bibnamefont {Mokhov}},
  \bibinfo {author} {\bibfnamefont {M.~V.}\ \bibnamefont {Muzafarova}},
  \bibinfo {author} {\bibfnamefont {S.~B.}\ \bibnamefont {Orlinskii}}, \ and\
  \bibinfo {author} {\bibfnamefont {J.}~\bibnamefont {Schmidt}},\ }\bibfield
  {title} {\enquote {\bibinfo {title} {{EPR} identification of the triplet
  ground state and photoinduced population inversion for a {S}i-{C} divacancy
  in silicon carbide},}\ }\href@noop {} {\bibfield  {journal} {\bibinfo
  {journal} {J. Exp. Theor. Phys. Lett.}\ }\textbf {\bibinfo {volume} {82}},\
  \bibinfo {pages} {441--443} (\bibinfo {year} {2005})}\BibitemShut {NoStop}%
\bibitem [{\citenamefont {Koehl}\ \emph {et~al.}(2011)\citenamefont {Koehl},
  \citenamefont {Buckley}, \citenamefont {Heremans}, \citenamefont {Calusine},\
  and\ \citenamefont {Awschalom}}]{Koehl2011}%
  \BibitemOpen
  \bibfield  {author} {\bibinfo {author} {\bibfnamefont {W.~F.}\ \bibnamefont
  {Koehl}}, \bibinfo {author} {\bibfnamefont {B.~B.}\ \bibnamefont {Buckley}},
  \bibinfo {author} {\bibfnamefont {J.~F.}\ \bibnamefont {Heremans}}, \bibinfo
  {author} {\bibfnamefont {G.}~\bibnamefont {Calusine}}, \ and\ \bibinfo
  {author} {\bibfnamefont {D.~D.}\ \bibnamefont {Awschalom}},\ }\bibfield
  {title} {\enquote {\bibinfo {title} {Room temperature coherent control of
  defect spin qubits in silicon carbide},}\ }\href@noop {} {\bibfield
  {journal} {\bibinfo  {journal} {Nature}\ }\textbf {\bibinfo {volume} {479}},\
  \bibinfo {pages} {84--87} (\bibinfo {year} {2011})}\BibitemShut {NoStop}%
\bibitem [{\citenamefont {Klimov}\ \emph {et~al.}(2015)\citenamefont {Klimov},
  \citenamefont {Falk}, \citenamefont {Christle}, \citenamefont {Dobrovitski},\
  and\ \citenamefont {Awschalom}}]{Klimov2015}%
  \BibitemOpen
  \bibfield  {author} {\bibinfo {author} {\bibfnamefont {P.~V.}\ \bibnamefont
  {Klimov}}, \bibinfo {author} {\bibfnamefont {A.~L.}\ \bibnamefont {Falk}},
  \bibinfo {author} {\bibfnamefont {D.~J.}\ \bibnamefont {Christle}}, \bibinfo
  {author} {\bibfnamefont {V.~V.}\ \bibnamefont {Dobrovitski}}, \ and\ \bibinfo
  {author} {\bibfnamefont {D.~D.}\ \bibnamefont {Awschalom}},\ }\bibfield
  {title} {\enquote {\bibinfo {title} {Quantum entanglement at ambient
  conditions in a macroscopic solid-state spin ensemble},}\ }\href@noop {}
  {\bibfield  {journal} {\bibinfo  {journal} {Sci. Adv.}\ }\textbf {\bibinfo
  {volume} {1}},\ \bibinfo {pages} {e1501015} (\bibinfo {year}
  {2015})}\BibitemShut {NoStop}%
\bibitem [{\citenamefont {Christle}\ \emph {et~al.}(2015)\citenamefont
  {Christle}, \citenamefont {Falk}, \citenamefont {Andrich}, \citenamefont
  {Klimov}, \citenamefont {Ul~Hassan}, \citenamefont {Son}, \citenamefont
  {Janz{\'e}n}, \citenamefont {Ohshima},\ and\ \citenamefont
  {Awschalom}}]{Christle2015}%
  \BibitemOpen
  \bibfield  {author} {\bibinfo {author} {\bibfnamefont {D.~J.}\ \bibnamefont
  {Christle}}, \bibinfo {author} {\bibfnamefont {A.~L.}\ \bibnamefont {Falk}},
  \bibinfo {author} {\bibfnamefont {P.}~\bibnamefont {Andrich}}, \bibinfo
  {author} {\bibfnamefont {P.~V.}\ \bibnamefont {Klimov}}, \bibinfo {author}
  {\bibfnamefont {J.}~\bibnamefont {Ul~Hassan}}, \bibinfo {author}
  {\bibfnamefont {N.~T.}\ \bibnamefont {Son}}, \bibinfo {author} {\bibfnamefont
  {E.}~\bibnamefont {Janz{\'e}n}}, \bibinfo {author} {\bibfnamefont
  {T.}~\bibnamefont {Ohshima}}, \ and\ \bibinfo {author} {\bibfnamefont
  {D.~D.}\ \bibnamefont {Awschalom}},\ }\bibfield  {title} {\enquote {\bibinfo
  {title} {Isolated electron spins in silicon carbide with millisecond
  coherence times},}\ }\href@noop {} {\bibfield  {journal} {\bibinfo  {journal}
  {Nat. mater.}\ }\textbf {\bibinfo {volume} {14}},\ \bibinfo {pages} {160}
  (\bibinfo {year} {2015})}\BibitemShut {NoStop}%
\bibitem [{\citenamefont {Anderson}\ \emph {et~al.}(2021)\citenamefont
  {Anderson}, \citenamefont {Glen}, \citenamefont {Zeledon}, \citenamefont
  {Bourassa}, \citenamefont {Jin}, \citenamefont {Zhu}, \citenamefont
  {Vorwerk}, \citenamefont {Crook}, \citenamefont {Abe}, \citenamefont
  {Ul-Hassan}, \citenamefont {Ohshima}, \citenamefont {Son}, \citenamefont
  {Galli},\ and\ \citenamefont {Awschalom}}]{5secSiC}%
  \BibitemOpen
  \bibfield  {author} {\bibinfo {author} {\bibfnamefont {C.~P.}\ \bibnamefont
  {Anderson}}, \bibinfo {author} {\bibfnamefont {E.~O.}\ \bibnamefont {Glen}},
  \bibinfo {author} {\bibfnamefont {C.}~\bibnamefont {Zeledon}}, \bibinfo
  {author} {\bibfnamefont {A.}~\bibnamefont {Bourassa}}, \bibinfo {author}
  {\bibfnamefont {Y.}~\bibnamefont {Jin}}, \bibinfo {author} {\bibfnamefont
  {Y.}~\bibnamefont {Zhu}}, \bibinfo {author} {\bibfnamefont {C.}~\bibnamefont
  {Vorwerk}}, \bibinfo {author} {\bibfnamefont {A.~L.}\ \bibnamefont {Crook}},
  \bibinfo {author} {\bibfnamefont {H.}~\bibnamefont {Abe}}, \bibinfo {author}
  {\bibfnamefont {J.}~\bibnamefont {Ul-Hassan}}, \bibinfo {author}
  {\bibfnamefont {T.}~\bibnamefont {Ohshima}}, \bibinfo {author} {\bibfnamefont
  {N.~T.}\ \bibnamefont {Son}}, \bibinfo {author} {\bibfnamefont
  {G.}~\bibnamefont {Galli}}, \ and\ \bibinfo {author} {\bibfnamefont {D.~D.}\
  \bibnamefont {Awschalom}},\ }\href@noop {} {\  (\bibinfo {year} {2021})},\
  \Eprint {http://arxiv.org/abs/arXiv:2110.01590} {arXiv:2110.01590}
  \BibitemShut {NoStop}%
\bibitem [{\citenamefont {\hbox{de las Casas}}\ \emph
  {et~al.}(2017)\citenamefont {\hbox{de las Casas}}, \citenamefont {Christle},
  \citenamefont {Ul~Hassan}, \citenamefont {Ohshima}, \citenamefont {Son},\
  and\ \citenamefont {Awschalom}}]{Casas2017}%
  \BibitemOpen
  \bibfield  {author} {\bibinfo {author} {\bibfnamefont {C.~F.}\ \bibnamefont
  {\hbox{de las Casas}}}, \bibinfo {author} {\bibfnamefont {D.~J.}\
  \bibnamefont {Christle}}, \bibinfo {author} {\bibfnamefont {J.}~\bibnamefont
  {Ul~Hassan}}, \bibinfo {author} {\bibfnamefont {T.}~\bibnamefont {Ohshima}},
  \bibinfo {author} {\bibfnamefont {N.~T.}\ \bibnamefont {Son}}, \ and\
  \bibinfo {author} {\bibfnamefont {D.~D.}\ \bibnamefont {Awschalom}},\
  }\bibfield  {title} {\enquote {\bibinfo {title} {Stark tuning and electrical
  charge state control of single divacancies in silicon carbide},}\ }\href@noop
  {} {\bibfield  {journal} {\bibinfo  {journal} {Appl. Phys. Lett.}\ }\textbf
  {\bibinfo {volume} {111}},\ \bibinfo {pages} {262403} (\bibinfo {year}
  {2017})}\BibitemShut {NoStop}%
\bibitem [{\citenamefont {Christle}\ \emph {et~al.}(2017)\citenamefont
  {Christle}, \citenamefont {Klimov}, \citenamefont {\hbox{de las Casas}},
  \citenamefont {Sz\'asz}, \citenamefont {Iv\'ady}, \citenamefont
  {Jokubavicius}, \citenamefont {Ul~Hassan}, \citenamefont {Syv\"aj\"arvi},
  \citenamefont {Koehl}, \citenamefont {Ohshima}, \citenamefont {Son},
  \citenamefont {Janz\'en}, \citenamefont {Gali},\ and\ \citenamefont
  {Awschalom}}]{Christle2017}%
  \BibitemOpen
  \bibfield  {author} {\bibinfo {author} {\bibfnamefont {D.~J.}\ \bibnamefont
  {Christle}}, \bibinfo {author} {\bibfnamefont {P.~V.}\ \bibnamefont
  {Klimov}}, \bibinfo {author} {\bibfnamefont {C.~F.}\ \bibnamefont {\hbox{de
  las Casas}}}, \bibinfo {author} {\bibfnamefont {K.}~\bibnamefont {Sz\'asz}},
  \bibinfo {author} {\bibfnamefont {V.}~\bibnamefont {Iv\'ady}}, \bibinfo
  {author} {\bibfnamefont {V.}~\bibnamefont {Jokubavicius}}, \bibinfo {author}
  {\bibfnamefont {J.}~\bibnamefont {Ul~Hassan}}, \bibinfo {author}
  {\bibfnamefont {M.}~\bibnamefont {Syv\"aj\"arvi}}, \bibinfo {author}
  {\bibfnamefont {W.~F.}\ \bibnamefont {Koehl}}, \bibinfo {author}
  {\bibfnamefont {T.}~\bibnamefont {Ohshima}}, \bibinfo {author} {\bibfnamefont
  {N.~T.}\ \bibnamefont {Son}}, \bibinfo {author} {\bibfnamefont
  {E.}~\bibnamefont {Janz\'en}}, \bibinfo {author} {\bibfnamefont {\'A.}\
  \bibnamefont {Gali}}, \ and\ \bibinfo {author} {\bibfnamefont {D.~D.}\
  \bibnamefont {Awschalom}},\ }\bibfield  {title} {\enquote {\bibinfo {title}
  {Isolated spin qubits in {S}i{C} with a high-fidelity infrared spin-to-photon
  interface},}\ }\href@noop {} {\bibfield  {journal} {\bibinfo  {journal}
  {Phys. Rev. X}\ }\textbf {\bibinfo {volume} {7}},\ \bibinfo {pages} {021046}
  (\bibinfo {year} {2017})}\BibitemShut {NoStop}%
\bibitem [{\citenamefont {Anderson}\ \emph {et~al.}(2019)\citenamefont
  {Anderson}, \citenamefont {Bourassa}, \citenamefont {Miao}, \citenamefont
  {Wolfowicz}, \citenamefont {Mintun}, \citenamefont {Crook}, \citenamefont
  {Abe}, \citenamefont {Hassan}, \citenamefont {Son}, \citenamefont {Ohshima},\
  and\ \citenamefont {Awschalom}}]{Anderson2019}%
  \BibitemOpen
  \bibfield  {author} {\bibinfo {author} {\bibfnamefont {C.~P.}\ \bibnamefont
  {Anderson}}, \bibinfo {author} {\bibfnamefont {A.}~\bibnamefont {Bourassa}},
  \bibinfo {author} {\bibfnamefont {K.~C.}\ \bibnamefont {Miao}}, \bibinfo
  {author} {\bibfnamefont {G.}~\bibnamefont {Wolfowicz}}, \bibinfo {author}
  {\bibfnamefont {P.~J.}\ \bibnamefont {Mintun}}, \bibinfo {author}
  {\bibfnamefont {A.~L.}\ \bibnamefont {Crook}}, \bibinfo {author}
  {\bibfnamefont {H.}~\bibnamefont {Abe}}, \bibinfo {author} {\bibfnamefont
  {J.~U.}\ \bibnamefont {Hassan}}, \bibinfo {author} {\bibfnamefont {N.~T.}\
  \bibnamefont {Son}}, \bibinfo {author} {\bibfnamefont {T.}~\bibnamefont
  {Ohshima}}, \ and\ \bibinfo {author} {\bibfnamefont {D.~D.}\ \bibnamefont
  {Awschalom}},\ }\bibfield  {title} {\enquote {\bibinfo {title} {Electrical
  and optical control of single spins integrated in scalable semiconductor
  devices},}\ }\href@noop {} {\bibfield  {journal} {\bibinfo  {journal}
  {Science}\ }\textbf {\bibinfo {volume} {366}},\ \bibinfo {pages} {1225--1230}
  (\bibinfo {year} {2019})}\BibitemShut {NoStop}%
\bibitem [{\citenamefont {Candido}\ and\ \citenamefont
  {Flatt\'e}(2021)}]{Candido2020}%
  \BibitemOpen
  \bibfield  {author} {\bibinfo {author} {\bibfnamefont {Denis~R.}\
  \bibnamefont {Candido}}\ and\ \bibinfo {author} {\bibfnamefont {Michael~E.}\
  \bibnamefont {Flatt\'e}},\ }\bibfield  {title} {\enquote {\bibinfo {title}
  {Suppression of the optical linewidth and spin decoherence of a quantum spin
  center in a $p$-$n$ diode},}\ }\href {\doibase 10.1103/PRXQuantum.2.040310}
  {\bibfield  {journal} {\bibinfo  {journal} {PRX Quantum}\ }\textbf {\bibinfo
  {volume} {2}},\ \bibinfo {pages} {040310} (\bibinfo {year}
  {2021})}\BibitemShut {NoStop}%
\bibitem [{\citenamefont {Song}\ and\ \citenamefont {Dery}(2014)}]{Song2014}%
  \BibitemOpen
  \bibfield  {author} {\bibinfo {author} {\bibfnamefont {Y.}~\bibnamefont
  {Song}}\ and\ \bibinfo {author} {\bibfnamefont {H.}~\bibnamefont {Dery}},\
  }\bibfield  {title} {\enquote {\bibinfo {title} {Magnetic-field-modulated
  resonant tunneling in ferromagnetic-insulator-nonmagnetic junctions},}\
  }\href@noop {} {\bibfield  {journal} {\bibinfo  {journal} {Phys. Rev. Lett.}\
  }\textbf {\bibinfo {volume} {113}},\ \bibinfo {pages} {047205} (\bibinfo
  {year} {2014})}\BibitemShut {NoStop}%
\bibitem [{\citenamefont {Yue}\ \emph {et~al.}(2015)\citenamefont {Yue},
  \citenamefont {Prestgard}, \citenamefont {Tiwari},\ and\ \citenamefont
  {Raikh}}]{Yue2015}%
  \BibitemOpen
  \bibfield  {author} {\bibinfo {author} {\bibfnamefont {Z.}~\bibnamefont
  {Yue}}, \bibinfo {author} {\bibfnamefont {M.~C.}\ \bibnamefont {Prestgard}},
  \bibinfo {author} {\bibfnamefont {A.}~\bibnamefont {Tiwari}}, \ and\ \bibinfo
  {author} {\bibfnamefont {M.~E.}\ \bibnamefont {Raikh}},\ }\bibfield  {title}
  {\enquote {\bibinfo {title} {Resonant magnetotunneling between normal and
  ferromagnetic electrodes in relation to the three-terminal spin transport},}\
  }\href@noop {} {\bibfield  {journal} {\bibinfo  {journal} {Phys. Rev. B}\
  }\textbf {\bibinfo {volume} {91}},\ \bibinfo {pages} {195316} (\bibinfo
  {year} {2015})}\BibitemShut {NoStop}%
\bibitem [{\citenamefont {Inoue}\ \emph {et~al.}(2015)\citenamefont {Inoue},
  \citenamefont {Swartz}, \citenamefont {Harmon}, \citenamefont {Tachikawa},
  \citenamefont {Hikita}, \citenamefont {Flatt\'e},\ and\ \citenamefont
  {Hwang}}]{Inoue2015}%
  \BibitemOpen
  \bibfield  {author} {\bibinfo {author} {\bibfnamefont {H.}~\bibnamefont
  {Inoue}}, \bibinfo {author} {\bibfnamefont {A.~G.}\ \bibnamefont {Swartz}},
  \bibinfo {author} {\bibfnamefont {N.~J.}\ \bibnamefont {Harmon}}, \bibinfo
  {author} {\bibfnamefont {T.}~\bibnamefont {Tachikawa}}, \bibinfo {author}
  {\bibfnamefont {Y.}~\bibnamefont {Hikita}}, \bibinfo {author} {\bibfnamefont
  {M.~E.}\ \bibnamefont {Flatt\'e}}, \ and\ \bibinfo {author} {\bibfnamefont
  {H.~Y.}\ \bibnamefont {Hwang}},\ }\bibfield  {title} {\enquote {\bibinfo
  {title} {Origin of the magnetoresistance in oxide tunnel junctions determined
  through electric polarization control of the interface},}\ }\href@noop {}
  {\bibfield  {journal} {\bibinfo  {journal} {Phys. Rev. X}\ }\textbf {\bibinfo
  {volume} {5}},\ \bibinfo {pages} {041023} (\bibinfo {year}
  {2015})}\BibitemShut {NoStop}%
\bibitem [{\citenamefont {Harmon}\ and\ \citenamefont
  {Flatt{\'e}}(2021)}]{Harmon2021}%
  \BibitemOpen
  \bibfield  {author} {\bibinfo {author} {\bibfnamefont {N.~J.}\ \bibnamefont
  {Harmon}}\ and\ \bibinfo {author} {\bibfnamefont {M.~E.}\ \bibnamefont
  {Flatt{\'e}}},\ }\bibfield  {title} {\enquote {\bibinfo {title} {Theory of
  oblique-field magnetoresistance from spin centers in three-terminal
  spintronic devices},}\ }\href@noop {} {\bibfield  {journal} {\bibinfo
  {journal} {Phys. Rev. B}\ }\textbf {\bibinfo {volume} {103}},\ \bibinfo
  {pages} {035310} (\bibinfo {year} {2021})}\BibitemShut {NoStop}%
\bibitem [{\citenamefont {McMillan}\ \emph {et~al.}(2020)\citenamefont
  {McMillan}, \citenamefont {Harmon},\ and\ \citenamefont
  {Flatt{\'e}}}]{McMillan2020}%
  \BibitemOpen
  \bibfield  {author} {\bibinfo {author} {\bibfnamefont {S.~R.}\ \bibnamefont
  {McMillan}}, \bibinfo {author} {\bibfnamefont {N.~J.}\ \bibnamefont
  {Harmon}}, \ and\ \bibinfo {author} {\bibfnamefont {M.~E.}\ \bibnamefont
  {Flatt{\'e}}},\ }\bibfield  {title} {\enquote {\bibinfo {title} {Image of
  dynamic local exchange interactions in the dc magnetoresistance of
  spin-polarized current through a dopant},}\ }\href@noop {} {\bibfield
  {journal} {\bibinfo  {journal} {Physical Review Letters}\ }\textbf {\bibinfo
  {volume} {125}},\ \bibinfo {pages} {257203} (\bibinfo {year}
  {2020})}\BibitemShut {NoStop}%
\bibitem [{\citenamefont {Haberkorn}(1976)}]{Haberkorn1976}%
  \BibitemOpen
  \bibfield  {author} {\bibinfo {author} {\bibfnamefont {R.}~\bibnamefont
  {Haberkorn}},\ }\bibfield  {title} {\enquote {\bibinfo {title} {Density
  matrix description of spin-selective radical pair reactions},}\ }\href@noop
  {} {\bibfield  {journal} {\bibinfo  {journal} {Mol. Phys.}\ }\textbf
  {\bibinfo {volume} {32}},\ \bibinfo {pages} {1491} (\bibinfo {year}
  {1976})}\BibitemShut {NoStop}%
\end{thebibliography}%

\end{document}